\documentstyle[epsfig,aps,prl,multicol]{revtex}
\begin{document}

\title{Electronic polarization at surfaces and thin films of organic
molecular crystals: PTCDA}

\author{E.V. Tsiper$^{(a)}$, Z.G. Soos$^{(a)}$, W. Gao$^{(b)}$, and
A. Kahn$^{(b)}$}

\address{$^{(a)}$Department of Chemistry, Princeton University,
Princeton, NJ 08544\\$^{(b)}$Department of Electrical Engineering,
Princeton University, Princeton, NJ 08544}

% \date{\today}
\date{May 2, 2002}

\maketitle

\begin{multicols}{2}
 [ \begin{abstract}
The electronic polarization energies, $P=P_++P_-$, of a PTCDA
(perylenetetracarboxylic acid dianhydride) cation and anion in a
crystalline thin film on a metallic substrate are computed and
compared with measurements of the PTCDA transport gap on gold and
silver.  Both experiments and theory show that $P$ is 500 meV larger
in a PTCDA monolayer than in 50 \AA\ films.  Electronic polarization
in systems with surfaces and interfaces are obtained self-consistently
in terms of charge redistribution within molecules.\\
 \end{abstract} ]

\section{Transport Gap}

The electronic structure of organic molecular crystals is strikingly
different from the conventional inorganic semiconductors, such as Si,
in that the electronic polarization of the dielectric medium by charge
carriers constitutes a major effect, with energy scale greater than
transfer integrals or temperature \cite{silinsh,pope}.  The transport
gap $E_t$ for creating a separated electron-hole pair has a
substantial (1---2 eV) polarization energy contribution
\cite{soos_cpl} and exceeds the optical gap by $\sim1$ eV.  Limited
overlap rationalizes the modest mobilities of organic molecular
solids.  Devices such as light-emitting diodes, thin film transistors,
or photovoltaic cells require charge transport and are consequently
based on thin films, quite often deposited on metallic substrates
\cite{forrest_review,schon_science_FET}.  Organic electronics relies
heavily on controlling films with monolayer precision, on forming
structures with several thin films, and on characterizing the
interfaces.  The positions of transport states and mechanisms for
charge injection are among the outstanding issues for exploiting
organic devices.  We focus here on the electronic polarization of
crystalline thin films near surfaces and interfaces.  We find that
electronic polarization, and hence $E_t$, in a prototypical organic
crystal is significantly different at a free surface, near a
metal-organic interface, in thin organic layers, and in the bulk.

Weak intermolecular forces characterize organic molecular crystals,
whose electronic and vibrational spectra are readily related to
gas-phase transitions \cite{silinsh,pope}.  Due to small transfer
integrals, charge carriers are molecular ions embedded in the lattice
of neutral molecules.  The transport gap $E_t$ in the crystal is
derived from the charge gap for electron transfer in the gas phase,
$I(g)-A(g)$, which is the difference between the ionization potential
and the electron affinity.  But crystals have electrostatic
interactions even in the limit of no overlap, and charge carriers are
surrounded by self-consistent polarization clouds.  In contrast to
polaronic effects, electronic polarization is instantaneous and
directly affects the positions of energy levels.  Formation of
polarization clouds is associated with stabilization energy $P_+$ for
cations (the ``holes'') and $P_-$ for anions (the ``electrons'').  The
combination $P=P_++P_-$ occurs in $E_t=I(g)-A(g)-P$.  Since Coulomb
interactions are long-ranged, polarization clouds extend over many
lattice constants and $P$ depends on the proximity to surfaces and
interfaces.

\vskip 0.1 in
\centerline{\epsfig{file=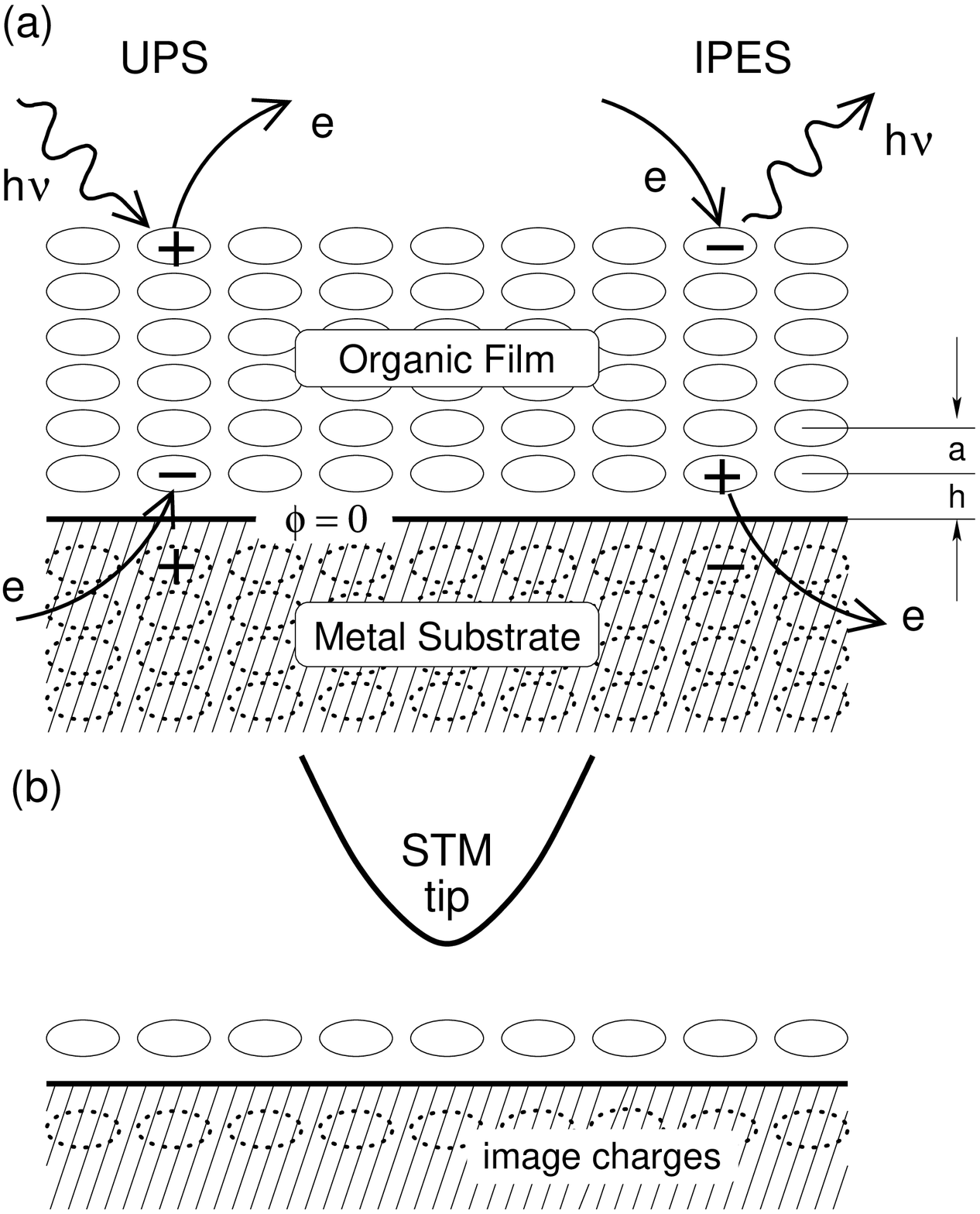,width=3 in}}

% \vskip 0.1 in

{\small {\bf Fig.~1} Schematic representation of charge-generation
processes in crystalline molecular films.  (a) UPS/IPES generates a
cation/anion at the outer surface, while charge injection from the
substrate involves the layer next to the metal.  (b) Tunneling through
a monolayer.  Dashed ovals in (a) and (b) represent image charges in
the metal.}

\vskip 0.1 in

Figure 1(a) depicts schematically an ultraviolet photoemission (UPS)
process where the ejected electron leaves behind a molecular cation in
the outermost layer.  In inverse photoelectron spectroscopy (IPES),
the surface is irradiated with low-energy electrons and the emitted
photon is detected when an electron is captured to form a molecular
anion.  UPS data is increasingly available from sub-monolayer to
$\sim$100\AA\ films \cite{seki}, while the IPES data is more limited.
The combination of UPS with IPES yields $E_t$ directly, with $P$ about
1---2 eV in representative organic materials used in devices
\cite{soos_cpl}.  As sketched in Fig.~1(b), tunneling electron
spectroscopy gives $E_t$ as the interval between the differential
conductance peaks when the potential of the tip matches either the
electron or the hole transport levels of the film.

\vskip 0.1 in
\centerline{\epsfig{file=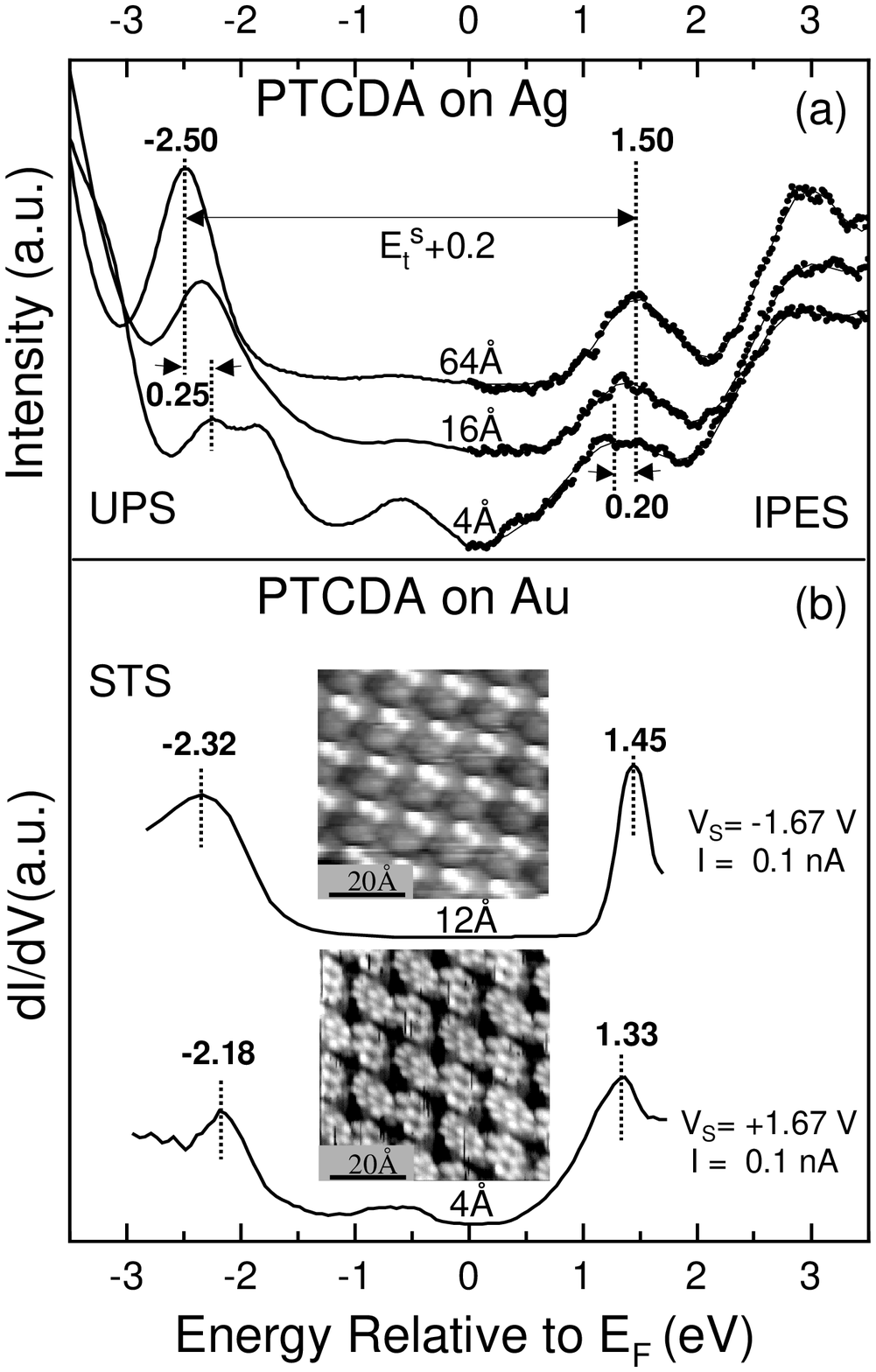,width=3 in}}

{\small {\bf Fig.~2} (a) Composite UPS/IPES spectra as a function of
PTCDA thickness on Ag.  Energy scales are aligned by measuring the
Fermi energy by UPS and IPES on Ag prior to PTCDA deposition.  (b)
dI/dV(V) STS spectra of filled and empty states recorded for a
monolayer (bottom) and a 2-3 molecular layer (top) film of PTCDA
deposited on Au.  The corresponding STM images of the films are
shown. The curves were recorded at the same tunneling setpoints as the
corresponding area scans.}

\vskip 0.1 in

Figure 2(a) shows UPS and IPES spectra of PTCDA
(perylenetetracarboxylic acid dianhydride) on silver.  PTCDA is an
excellent former of crystalline films whose structures are close,
though not identical, to having a (102) plane of the bulk crystal in
contact with the substrate \cite{forrest_review}.  The measured
transport gap on thick ($>50$\AA) films is $E_t^{\rm S}=3.8$ eV on Ag,
in excellent agreement with PTCDA on Au \cite{soos_cpl}.  We use PTCDA
on Ag rather than Au because the Au(5d) levels interfere with UPS of a
monolayer.  $E_t^{\rm S}$ includes a 0.2 eV intramolecular vibrational
contribution \cite{soos_cpl} that reduces the UPS/IPES gap.  This
correction is the same for all PTCDA films.  Careful analysis of peak
positions indicates $\sim$200 meV shifts of both the cation level
(UPS) and anion level (IPES) between mono- and multilayer films.

Figure 2(b) shows the results of scanning tunneling spectroscopy (STS)
on a monolayer (lower spectrum) and a 2-3 layer film (upper spectrum)
of PTCDA on Au(111).  These dI/dV(V) spectra represent the density of
filled and empty states involved in tunneling out of and into the
layer, respectively.  Each spectrum is the average of 25 spectra
recorded at various points on highly ordered molecular layers.
High-resolution scanning tunneling microscopy (STM) images of these
layers, taken concomitantly with STS, show the characteristic
difference \cite{chizhov} between monolayers, in which molecules are in
contact with, and parallel to, the Au surface and appear symmetric,
and second and subsequent layers, in which the tilt angle of molecules
introduces an asymmetry in their STM image.

The tunneling spectrum of the monolayer shows peaks leading to
$E_t^{\rm ML}=3.3$ eV for the energy difference between adding an
electron or hole.  Remarkably, this energy difference increases by
about 0.25 eV on the 2-3 layer spectrum, in excellent agreement with
the intermediate UPS/IPES spectra of a 16\AA\ film (Fig.~2a).  Each
peak shifts away from the Fermi level by a roughly equal amount with
increasing coverage.  We note that $E_t^{\rm S}$ and $E_t^{\rm
S}-E_t^{\rm ML}$ show no dependence on dipoles at the metal-organic
interface, which have opposite signs for PTCDA on Ag and Au
\cite{hill_dipole}.  The dipole at the PTCDA/Ag interface corresponds
to electron transfer from the metal to interface molecules, which
gives rise to the filled gap states at $-0.6$ and $-1.8$ eV on the
4\AA\ UPS spectrum.  The broad feature around $-0.6$ eV on the 4\AA\
PTCDA/Au STS spectrum corresponds to the Au surface state, which is
not eliminated by the deposition of organic molecules \cite{pflaum}.

Reduced polarization energy at surfaces has long been appreciated on
general grounds \cite{silinsh,pope}.  As anticipated and found for
thin films of anthracene on Au, $P_+$ is about 200 meV smaller in the
surface layer \cite{salaneck_prl,duke}.  With similar reductions
expected for $P_-$, the transport gap at the surface is {\em
increased} some 400 meV from its bulk value, $E_t$.  The gap increases
near surfaces because vacuum is not polarizable.  Conversely, $E_t$
decreases near organic-metal interfaces due to the high polarizability
of metals.  These opposite contributions partially cancel in very thin
films on metal substrates.

\section{Charge Redistribution}

We note that $E_t$, which has been the focus of theoretical study
\cite{munn,silinsh} and governs bulk transport, is beyond the reach of
the surface experimental techniques that access the outermost layers
of the material and thus reflect $E_t^{\rm S}$.  Similarly, the
transport gap $E_t^{\rm M}$ of the layer next to the metal is relevant
for charge injection (Fig.~1(a)).  A gap reduction of several hundred
meV is significant since it directly alters interface barriers.

No theoretical treatment exists for the electronic polarization near
surfaces and interfaces, which require an accuracy of $\sim$100 meV.
Methods to estimate $P_\pm$ in the bulk have been developed, primarily
for the acenes, based on the microelectrostatics of polarizable points
that represent organic molecules \cite{munn,petelenz}.  Dielectric
response of a neutral surface has also been studied
\cite{munn_surface2000}.  We have recently developed an approach based
on the analysis of charge redistribution in organic molecules
\cite{ours_cpl}, and demonstrated accurate calculations of $P_+$,
$P_-$, and energies of ion pairs in bulk PTCDA and anthracene crystals
\cite{ours_prb}.  We apply here the same approach to calculate
polarization in thin organic films.

In the zero-overlap limit, molecules comprising the crystal are
quantum-mechanical objects interacting by classical forces.  A
self-consistent problem can be formulated \cite{ours_prb} that treats
molecules rigorously in the external fields of all other molecules.
We describe charge redistribution in organic molecules in terms of the
atom-atom polarizability tensor $\Pi_{ij}$ that relates a change in
the partial charge at an atom $i$ due to the electrostatic potential
$\phi_j=\phi({\bf r}_j)$ at atom $j$:

\begin{equation}
\rho_i=\rho_i^{(0)}-\sum_j\Pi_{ij}\phi_j
\label{Pi}
\end{equation}
 $\rho_i^{(0)}$ are the atomic charges in an isolated molecule.  The
tensor $\Pi_{ij}$ is a natural extension of the similar quantity in
$\pi$-electron theory \cite{coulson}.  We compute $\Pi_{ij}$ using
INDO/S \cite{indos}, which is a semiempirical Hamiltonian designed for
spectroscopic molecular properties.  $\Pi_{ij}$ describes the major,
``charge-induced'' part of molecular polarizability $\alpha^{\rm C}$,
which is augmented to reflect the actual polarizability $\alpha$ by
introducing induced atomic dipoles $\bbox{\mu}_i$ and distributing the
difference $\widetilde\alpha=\alpha-\alpha^{\rm C}$ over 38 atoms of
PTCDA in the spirit of submolecular methods \cite{munn}.
Self-consistent equations for $\rho_i$ and $\bbox{\mu}_i$ are then
solved for increasing cluster sizes of mesoscopic dimensions
($\sim$100\AA), and the macroscopic limits are found.  For neutral
lattices the approach has yielded accurate anisotropic dielectric
tensors of two representative organic crystals \cite{ours_cpl}.

We use identical molecular inputs here to model PTCDA films as
infinite slabs terminated by (102) planes next to a vacuum and a
metal, respectively, as sketched in Fig.~1.  The metal is taken as a
constant-potential plane at $z=0$, a distance $h$ from the innermost
molecular layer.  Image charges and dipoles at $z<0$ ensure that the
potential $\phi(z=0)=0$.  Any $\phi(0)=C$ leads to the same result for
$P=P_++P_-$, since one charge is stabilized and the other is
destabilized.  The metal-organic separation $h$ is a model parameter
that is related to Van der Waals radii.

\vskip 0.1 in
\centerline{\epsfig{file=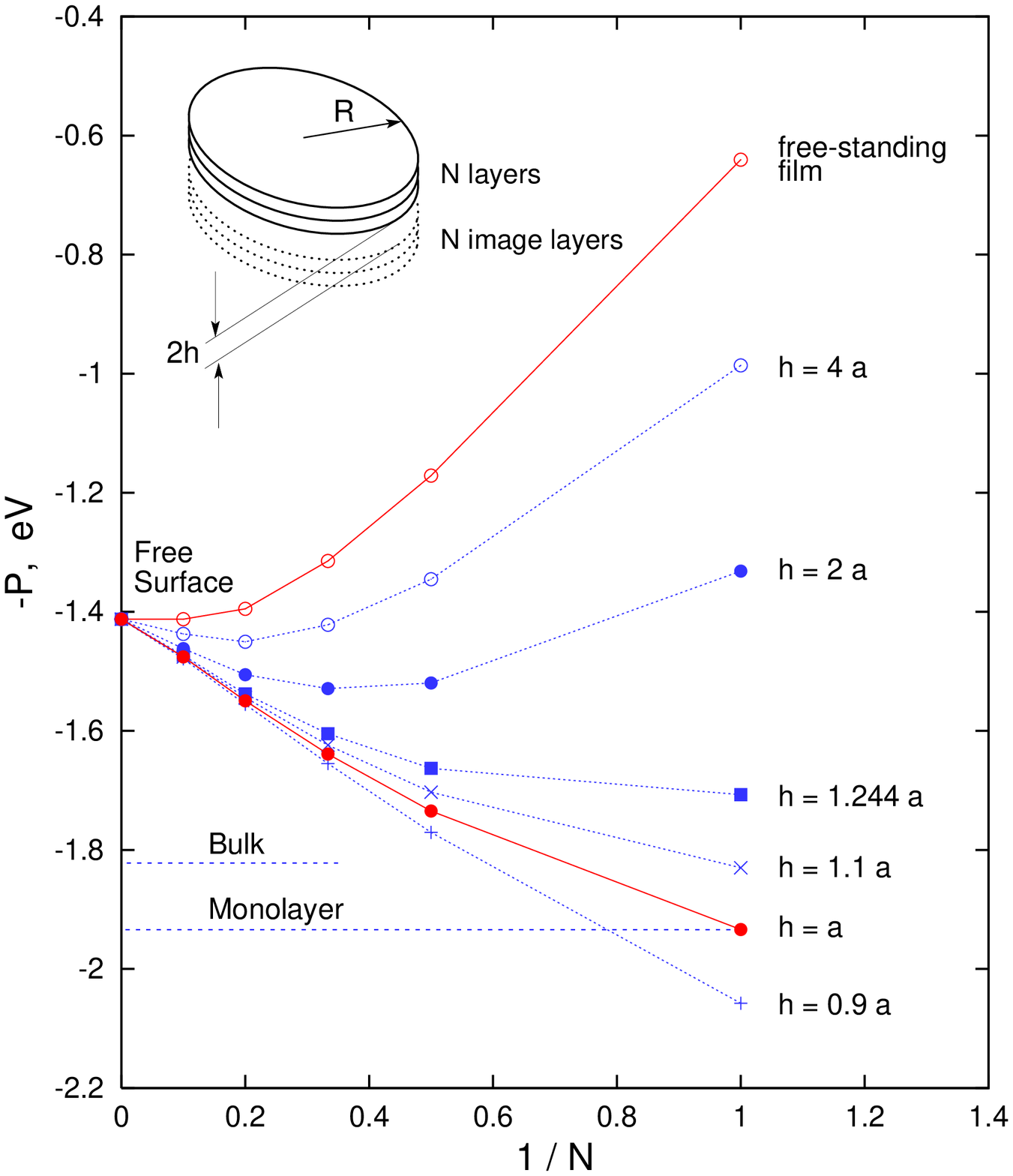,width=3 in}}

{\small {\bf Fig.~3} Electronic polarization $P=P_++P_-$ at the outer
layer of an $N$-layer film (Fig.~1) at separation $h$ from the metal.
The $P$ values at the free surface ($N\rightarrow\infty$), a monolayer
($N=1$), free-standing films ($h\rightarrow\infty$), and the bulk are
indicated.  A pill-box geometry with $R\sim100$\AA\ is used.  The
inset shows the transport gap, $E_t^{(n)}$, across a 10-layer film
with $n=1$ next to the metal and $n=10$ at the outer surface.}

\vskip 0.1 in

We use a pill-box geometry, defined by the radius $R$ (sketched in
Fig.~3) with the ion placed in one of the layers near the center of
the box.  We initially find self-consistent atomic charges and induced
dipoles $\overline\rho_i^k$ and $\overline{\bbox{\mu}}_i^k$,
$k=1,...,N$ in a neutral film of $N$ layers, where translational
symmetry gives rapid convergence.  We then consider pill-boxes of
increasing diameter $2R$ to find self-consistent
$\rho_i^a-\overline\rho_i^k$ and
$\bbox{\mu}_i^a-\overline{\bbox{\mu}}_i^k$ for every molecule $a$ in
the pill-box.  The largest systems ($2R=135$ \AA\ , $N=10$) contain
$\sim$2400 PTCDA molecules and their images, and larger values of $R$
can be used.  $P_+$ is the energy difference between a neutral film
and one containing a cation.  $P_-$ is the corresponding energy for an
anion in the same position.  We note that the energy of an infinite
film is extensive, but the difference is finite and can be evaluated
\cite{ours_prb}.  All data reported below is in the limit
$R\rightarrow\infty$.  $P$ thus depends on which layer contains the
ion.

\section{Results}

For the ion in the outermost molecular layer in the limit
$N\rightarrow\infty$ we obtain the free (102) surface polarization
$P^{\rm S}=1.41$ eV.  This limit does not depend on $h$ or the metal.
$P^{\rm S}$ is 0.41 eV less than the bulk value $P=1.82$ eV
\cite{ours_prb}.  The difference is consistent with experimental
estimates \cite{salaneck_prl} and corresponds to the surface
correction $c=1-P^{\rm S}/P=0.23$, where the value inferred
\cite{soos_cpl} from UPS and IPES spectra was $c\sim0.25$.

Figure 3 gives results for PTCDA layers of finite thickness.  Unlike
bulk or free-surface calculations, which are essentially
parameter-free, the finite layer data depends on $h$, which is the
only way the metal enters our idealized model.  The reasonable value
$h=a$ places the metal plane one lattice spacing $a=3.214$ \AA\ from
the innermost layer, which is also consistent with Van der Waals
radii, and 10 \% variations of $h$ are shown in Fig.~3.  We also
computed $P$ at $h=1.244a$, $2a$ and $4a$, and extrapolated as $1/h$.
The limit $h\rightarrow\infty$ gives the polarization at the surface
of a free-standing film of $N$ layers.

All curves in Fig.~3 converge to the free-surface value $P^{\rm
S}=1.41$ eV.  Using the curve $h=a$ we find the single monolayer value
$P^{\rm ML}=1.93$ eV, which corresponds to the tunneling spectroscopy
setup (Fig.~1(b)).  We see that for a monolayer on the metal surface,
the polarization energy is indeed close to the bulk value in line with
the expected cancellation discussed above.  The difference $P^{\rm
ML}-P^{\rm S}=0.52$ eV agrees with the experimental $E_t^{\rm
S}-E_t^{\rm ML}=0.45$---0.50 eV for PTCDA on Ag or Au.  In fact, the
agreement is slightly better since UPS/IPES data is for the the films
of finite thickness $N\sim20$.  The free-standing monolayer has
$P=0.64$ eV, which is about $E_t/3$.  Such a big polarization is
consistent with large in-plane polarizability of PTCDA molecules.
These results for $P$ are summarized for comparison in Table I.

\begin{center}

 \vskip 0.05 in
 {\small {\bf Table I}.\ \ \ \ $P_++P_-$ at various positions in the
film}
 \vskip 0.05 in

\begin{tabular}{lr}
% \\
\tableline
\tableline
% Position of the charge carrier & $(P_++P_-)$ \\
% \tableline
Bulk & 1.82 eV \\
Free surface & 1.41 eV \\
Monolayer, free-standing & 0.64 eV \\
Monolayer on metal ($h=a$) & 1.93 eV \\
Surface of a bilayer on metal ($h=a$) & 1.73 eV \\
Layer next to metal, thick film ($h=a$)\ \ \ \  & 2.21 eV
% Single molecule on metal & 1.83 eV
\\
\tableline
\tableline
% \multicolumn{4}{l}{$^a$ Eq.~(\protect\ref{Vapprox})}
\\
\end{tabular}

\end{center}

Analysis of UPS data for films of the electron-transport molecule
Alq$_3$ [tris(8-hydroxy-quinoline)-aluminum] on
silver \cite{hill_bending2000} yields strikingly similar conclusions:
the transport gap for a monolayer on metal substrate, $E_t^{\rm ML}$,
is equal to the inferred bulk value, and is about 400 meV narrower
than the gap in the outermost surface layer of a thick film, $E_t^{\rm
S}$.  Also, the inferred gap in the innermost layer of a thick film on
metal surface, $E_t^{\rm M}$, is about 400 meV
less \cite{hill_bending2000} than $E_t^{\rm ML}$.

% \vskip 0.1 in
\centerline{\epsfig{file=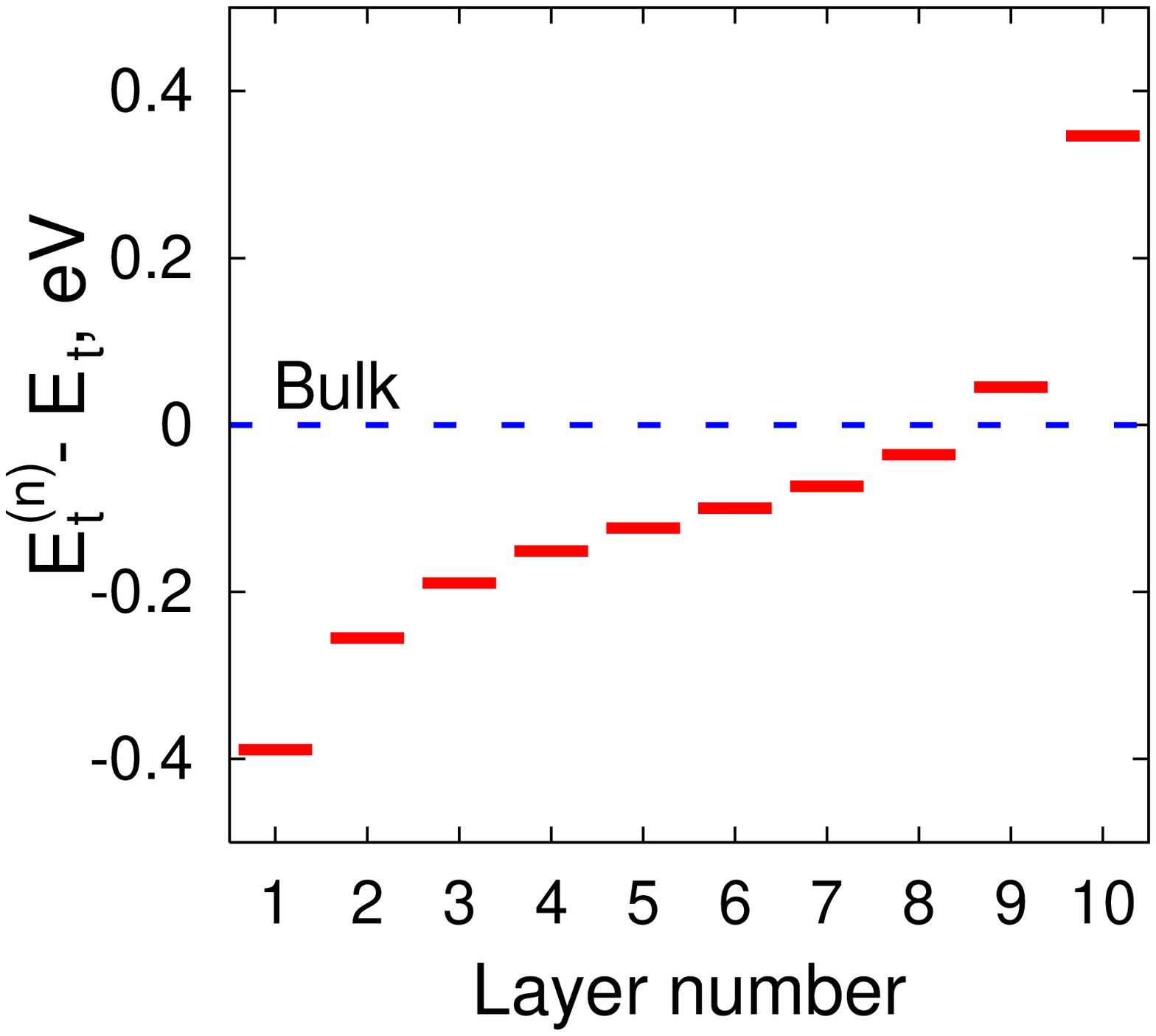,width=3 in}}

{\small {\bf Fig.~4} Variation of the transport gap from the bulk
value, $E_t^{(n)}-E_t$, across a 10-layer film with $n=1$ next to the
metal and $n=10$ at the outer surface.}

\vskip 0.1 in

Figure 4 shows the variation of the transport gap $E_t$ across a
10-layer PTCDA film on a metal substrate ($h=a$).  We note that
surface effects extend several layers into the sample.  The long-range
nature of surface polarization has been largely neglected.  An
influential early UPS study of 20\AA\ vapor-deposited anthracene films
on Au ascribed the 200 meV shift of $P_+$ to the single outermost
layer \cite{salaneck_prl}.  The additivity of polarization
contributions suggested \cite{hill_bending2000} for Alq$_3$ is tacitly
based on short-range interactions.  Greater polarizability next to the
metal is consistent with strong charge confinement to interfaces, as
inferred recently for pentacene field-effect transistors with
remarkable electronic characteristics \cite{schon_science_FET}.  In
general, the 400 meV increase of $P_++P_-$ at the metal-organic
interface is not shared equally by the electron and hole.  The
stabilization of either carrier by roughly 200 meV is important for
matching transport levels in injection.

Weak overlap in molecular solids constitutes, in fact, a significant
simplification over covalent bonding in inorganic semiconductors.
Even though charge transport is critical for electronic applications,
the picture of localized carriers is the proper zeroth-order
approximation, to which the overlap (i.e. kinetic energy) should be
considered as a perturbation.  This does not by itself rule out the
band-like description.  Rather, charged quasiparticles are to be
understood as surrounded by polarization clouds, likely affecting the
bandwidths.

Our results for $P_+$ and $P_-$ are exclusively electronic.  Lattice
relaxation around charges are considered to be small corrections
($\sim$10\%) on general grounds \cite{silinsh,pope}.  The idealized
model of image charges does not depend on the metal's Fermi energy or
on surface dipoles, whose shifts cancel in $P=P_++P_-$.  It also
ignores surface states or surface reactions that are known to occur at
specific organic/metal inferfaces \cite{seki}.  Since the
self-consistent calculation requires the film's structure, it is not
directly applicable to amorphous or structurally uncharacterized
films.

Support of this work by the NSF (DMR-0097133) is gratefully
acknowledged.

\end{multicols}
\end{document}